\documentclass[amsmath,amssymb,aps,prd,data,nofootinbib]{revtex4}
\usepackage[utf8]{inputenc}
\usepackage{graphicx}
\usepackage{amsfonts}
\usepackage{amssymb}
\usepackage{amsmath}
\usepackage{float}
\usepackage{dcolumn}
\usepackage{bm}
\usepackage{latexsym,color}

\begin{document}

\title{Equilibrium Configurations of Rotating White Dwarfs at Finite Temperatures}

\author{Kuantay Boshkayev\\
\small\itshape NNLOT, al-Farabi Kazakh National University,\\
\small\itshape al-Farabi ave. 71, 050040, Almaty, Kazakhstan.\\
\small\itshape ICRANet, Piazza della Repubblica 10, I--65122 Pescara, Italy.\\
\small\itshape kuantay@mail.ru}

\begin{abstract}
In this work, cold and hot, static and rotating white dwarf stars are
investigated within the framework of classical physics, employing the
Chandrasekhar equation of state. The main parameters of white dwarfs such as
the central density, pressure, total mass and radius are calculated fulfilling
the stability criteria for hot rotating stars. To construct rotating
configurations the Hartle approach is involved. It is shown that the effects of
finite temperatures become crucial in low-mass white dwarfs, whereas rotation
is relevant in all mass range. The simultaneous accounting for temperature and
rotation is critical in the calculation of the radii of white dwarfs. The
results obtained in this work can be applied to explain a variety of
observational data for white dwarfs from the Sloan Digital Sky Survey Data
Releases.
\end{abstract}

\maketitle

\section{Introduction}
Compact objects are the end products of stellar evolution and they are
subdivided into the basic three categories: white dwarfs (WDs), neutron stars
(NSs) and black holes (BHs), with the exception of exotic and at the same time
hypothetical objects such as quark stars, boson stars, gravastars, etc
\cite{zelnov71,shapirobook,glen2000,haenselbook}. These objects are called
compact because of their large mass and small size and, correspondingly, high
density. It is believed that the initial mass is a key factor determining the
final fate of a star. For example, WDs are formed from low-mass star
progenitors with masses $M\approx(1-8)M_{\odot}$ (solar mass) \cite{koester83},
though the lower and upper bounds of the progenitor mass are not well
constrained both from theory and observations \cite{berro2008}. Nonetheless,
the upper limit of the mass of a static cold WD without a magnetic field does
not exceed the Chandrasekhar mass limit $M\leq1.44M$ \cite{chandra31}.

The ratio of the gravitational radius to the actual radius of an object, the
so-called compactness parameter $r_g/R$, for WDs is $\sim0.001$, for NSs is
$\sim0.3$, for BHs is equal to $1$ \cite{shapirobook}, where $r_g=2GM/c^2$ is
the gravitational (Schwarzschild) radius, $G$  is the gravitational constant,
$M$ is the total mass of the object, $c$ is the speed of light in vacuum. From
here it is evident that the role of general relativity (GR) becomes more
pronounced when the compactness parameter increases. The importance of GR in
the case of massive WDs is well-known in the literature \cite
{zelnov71,shapirobook}. In accordance with Ref. \cite{sedrakyan71} and \cite{rrrx},
it is necessary to investigate WDs in GR in order to analyze their stability
against the relativistic corrections and small perturbations, though they can
be neglected for low-mass WDs.

According to the latest observational data by 2017 there are more than 32 000
registered WDs \cite{keplerscience}, which are splitted into groups and
subgroups depending on their mass, temperature, nuclear composition, magnetic
field and other physical characteristics. The data are available online and are
provided with the description and technical details of observations
\cite{kepler2,kepler1,koester2}.

In general, WDs are crucial to understand the accelerated expansion of the
universe in terms of type Ia supernova explosions, they can provide independent
information about the age of our galaxy and their distribution contains
evidences about star formation history and subsequent evolution. The
progenitors of  WDs  evolve and age on the stage of the main sequence star
losing carbon, nitrogen, oxygen etc. For this very reason they supply a
substantial input to the chemical evolution of our Galaxy and possibly they can
be considered a key source of life supporting chemical compounds
\cite{kepler2007}.

Currently, there are three major equations of state (EoSs) for describing the
degenerate matter of WDs: the classical Chandrasekhar EoS, the Salpeter EoS,
and the relativistic Feynman-Metropolis-Teller (RFMT) EoS. The RFMT EoS
generalizes the well-known Chandrasekhar and Salpeter EoSs, including the
effects of the Coulomb interactions and the local inhomogeneities of the
electron distribution within a full relativistic fashion. As a result, the
masses of WDs are smaller and the radii are larger than those obtained from the
Chandrasekhar's and Salpeter's EoSs. The principal differences, advantages and
drawbacks among these EoSs are amply described in Ref.~\cite{rrrx}. It
should also be noted that the polytropic EoSs, widely used in the literature,
are only the limiting cases of the Chandrasekhar or Salpeter EoSs in the
non-relativistic and extremely relativistic limits \cite{zelnov71,shapirobook}.

Throughout the paper WDs are studied using the Chandrasekhar EoS
\cite{b2016b,b2016c} at finite-temperatures in classical physics for the sake
clarity and simplicity. A similar approach of the inclusion of
finite-temperature effects in the RFMT EoS was analyzed in
Ref.~\cite{carvalho2014}. The main goal of the paper is to investigate the
influence of both rotation and finite-temperatures on the structure of WDs.
Accounting for such effects makes the theory of WDs be more realistic and
practical \cite{boshijmpe, brrs2013, bjkps2014, bmg13, b2016b, b2016c}.

\section{The Chandrasekhar equation of state at zero temperature}

The EoS of degenerate WD matter, in the simplest case, determines the
dependence of the total pressure on the total energy density. The substance of
WDs consists of electrons and positively charged ions (naked nuclei). The
electrons are considered as a fully degenerate electron gas and they are
described by the Fermi-Dirac statistics \cite{landaustat}. In the Chandrasekhar
approximation, the distribution of electrons, as well as ions, is assumed to be
locally constant \cite{rrrx}. Consequently, the condition of local charge
neutrality is given by
\begin{equation}\label{eq:1}
n_e=\frac{Z}{A}n_N,
\end{equation}
where $n_e$ is the number density of electrons, $Z$ is the number of protons,
$A$  is the average atomic weight (mass number), $n_N$ is the number density of
nucleons. In a fully degenerate case, all lower energy levels are filled up to
some maximum level, called the Fermi level. The number density of the fully
degenerate electron gas up to the Fermi level is defined as
\begin{equation}\label{eq:2}
n_e=\int_{0}^{p_e^F}\frac{2}{(2\pi\hbar)^3}d^3p=\frac{8\pi}{(2\pi\hbar)^3}\int_{0}^{p_e^F}p^2dp=\frac{(p_e^F)^3}{3\pi^2\hbar^3},
\end{equation}
where $p_e^F$  is the Fermi momentum of an electron, $\hbar$  is the reduced
Planck constant. According to the Chandrasekhar approximation the resulting
pressure is due to the electron pressure $P_e$, while the pressure of
positively charged nuclei $P_N$  is insignificant, and the energy density is
determined by the energy density of nuclei ${\cal E}_N$, while the energy
density   of degenerate electrons ${\cal E}_e$ is negligibly small. Thus, the
Chandrasekhar EoS is defined as \cite{chandra31}
\begin{eqnarray}\label{eq:3}
{\cal E}_{Ch}={\cal E}_N+{\cal E}_e\approx{\cal E}_N,\\\nonumber
P_{Ch}=P_N+P_e\approx P_e,
\end{eqnarray}
The resulting energy of nucleons by definition is given as
\begin{equation}\label{eq:4}
{\cal E}_N=\frac{A}{Z}M_u c^2n_e,
\end{equation}
where $M_u=1.66604\times10^{-24}$g is the unified atomic mass unit. The ratio
of the atomic number to the number of protons is usually denoted in the
literature as $\mu=A/Z$ and all calculations in this paper were carried out by
adopting $\mu=2$ for simplicity. The total pressure of electrons is defined as
\begin{eqnarray}\label{eq:5}
P_e=\frac{1}{3}\frac{2}{(2\pi\hbar)^3}\int_{0}^{p_e^F}\frac{c^2p^2}{\sqrt{c^2p^2+m_e^2c^4}}4\pi p^2dp \quad\qquad\qquad\qquad \\
\quad=\frac{m_e^4c^5}{8\pi^2\hbar^3}\left[x_e\sqrt{1+x_e^2}(2x_e^2/3-1)+\ln(x_e+\sqrt{1+x_e^2})\right], \nonumber
\end{eqnarray}
where $x_e=p_e^F/(m_e c)$ is the dimensionless Fermi momentum and $m_e$ is the electron mass \cite{rrrx}.

\section{The Chandrasekhar equation of state at finite temperatures}

In general, the expression for the electron number density follows from the
Fermi-Dirac statistics and, when temperature is taken into account, it is
determined as
\begin{equation}\label{eq:6}
n_e=\frac{2}{(2\pi\hbar)^3}\int_{0}^{\infty}\frac{4\pi p^2 dp}{\exp\left[\frac{E(p)-\mu_e(p)}{k_B T}\right]+1},
\end{equation}
where $k_B=1.38\times10^{-16}$erg K$^{-1}$ is the Boltzmann constant,
$T$ is the temperature, $\mu_e$  is the chemical potential,
$E(p)=\sqrt{c^2p^2+m_e^2c^4}-m_ec^2$ is the kinetic energy, $p$  and $m_e$  are
the momentum and the rest mass of an electron, respectively.

\begin{figure}
\centerline{\includegraphics[width=8.5cm]{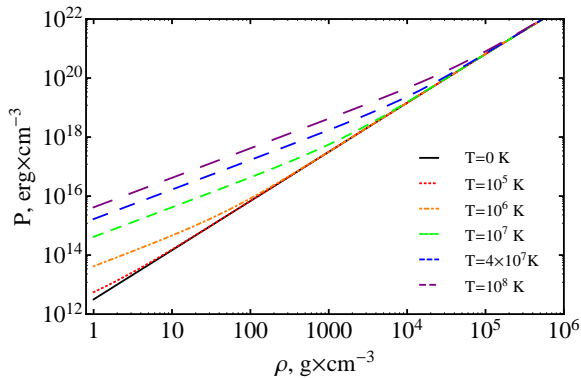}}
\caption{Total pressure as a function of the mass density for selected temperatures in the range $T=\left(0-10^8\right)$K (colour online).}
\label{fig:Prho}
\end{figure}

Formula (\ref{eq:2}), taking into account the effects of finite temperatures,
can be written in the following alternative form
\begin{equation}\label{eq:7}
n_{e}=\frac{8\pi \sqrt{2}}{(2 \pi \hbar)^3} m^3 c^3 \beta^{3/2} \left[ F_{1/2} (\eta,\beta) + \beta F_{3/2} (\eta,\beta) \right],
\end{equation}
where
\begin{equation}
F_{k} (\eta,\beta)=\int_{0} ^{\infty} \frac{t^k \sqrt{1+(\beta/2)t}}{1+ e^{t-\eta}} dt
\label{eq:Fk1}
\end{equation}
is the relativistic Fermi-Dirac integral, $\eta=\mu_e/(k_B T)$,
$t=E(p)/(k_{B}T)$ and $\beta=k_{B}T/(m_e c^2)$ are the degeneracy parameters
\cite{carvalho2014,timmes99}. Consequently, the total electron pressure for
$T\neq0$ K is given by
\begin{align}\label{eq:8}
P_e=\frac{2^{3/2}}{3 \pi^2 \hbar^3} m_{e}^4 c^5 \beta^{5/2} \left[ F_{3/2} (\eta,\beta)+ \frac{\beta}{2} F_{5/2} (\eta,\beta) \right].
\end{align}
The dependence of the total pressure on the total density Eq.~(\ref{eq:3}) at
various temperatures $T = (0, 10^5, 10^6, 10^7, 10^8)$ K is plotted in
Fig.~\ref{fig:Prho}. As one can see, the effects of temperature become
noticeable only at lower densities starting from $10^5$ g cm$^{-3}$. For
higher densities the thermal effects are negligible.

\section{Formalism and stability criteria for rotating white dwarfs at finite temperatures}

It has been established that for WDs relativistic effects lead only to small
perturbations of Newtonian gravity \cite{mn14}. Consequently, Newton's theory
in the low mass region allows one to study sufficiently well the essential
physical features of WDs. We use the classical limit of the Hartle-Thorne
formalism \cite{Hartle, Hartle2} to analyze  perturbatively the structural
equations \cite{bosh2016}. The basic idea consists in solving Newton's field
equation
\begin{equation}
\nabla^2 \Phi = 4 \pi G \rho\ ,
\label{field}
\end{equation}
and the structure equations
\begin{equation}
\frac{d P }{dr } = - \rho \frac{G M}{r^2} \ ,\qquad \frac{dM}{dr} = 4 \pi r^2 \rho\ ,
\label{equi}
\end{equation}
perturbatively by expanding the radial coordinate as $r=R + \xi$. The structure
equations contain the hydrostatic equilibrium condition between gravitational
and pressure forces, and the mass balance equation. Hence, here $\Phi$ is the
gravitational potential, $\rho$ is the matter density related to the energy
density as ${\cal E}=c^2\rho$, $P$ is the pressure, $M(r)$ is the mass
inside a sphere with radius $r$, $R$ is the radial coordinate for a spherical
configuration and the function $\xi(R,\theta)$ takes into account the
deviations from spherical symmetry due to the rotation of the star.

All the important quantities such as the total mass $M$, equatorial radius
$R_e$, moment of inertia $I$, angular momentum $J$, quadrupole moment $Q$, etc.
are then Taylor expanded up to the second order in the angular velocity. Within
the Hartle approach, due to a proper choice of function $\xi$, the
density $\rho$ and pressure $P$ can be treated as non affected by the rotation
of the star. The field and structural equations (\ref{field}) and (\ref{equi})
can then be integrated numerically to obtain all the important quantities in
the preferred approximation \cite{bosh2016}.

For our analysis it is convenient to introduce the
Keplerian angular velocity
\begin{equation}\label{eq:omegaK}
\Omega_{Kep}=\sqrt{\frac{GM}{R^3_e}} \ ,
\end{equation}
because it allows us to calculate all the fundamental parameters at the
mass-shedding limit, and to determine the stability region inside which
rotating configurations can exist \citep{brrs2013}.

Finally, the inverse $\beta$-decay instability determines the critical density
which, in turn, defines the onset of instability for a WD to collapse into a
NS. Thus the inverse $\beta$-decay instability is crucial both for static and
rotating configurations. It represents one of the boundaries of the stability
region for rotating WDs \citep{brrs2013, bosh2016}. According to
Ref.~\cite{carvalho2014}, the occurrence of the inverse $\beta$-decay
instability is not affected by the presence of temperature, i. e. it is the
same as in the Chandrasekhar EoS $\rho_{crit}=1.37\times10^{11}$
g$\times$cm$^{-3}$. This  is related to the fact that the effects of
temperature are negligible in the higher density regime.

\begin{figure}
\centerline{\includegraphics[width=8.5cm]{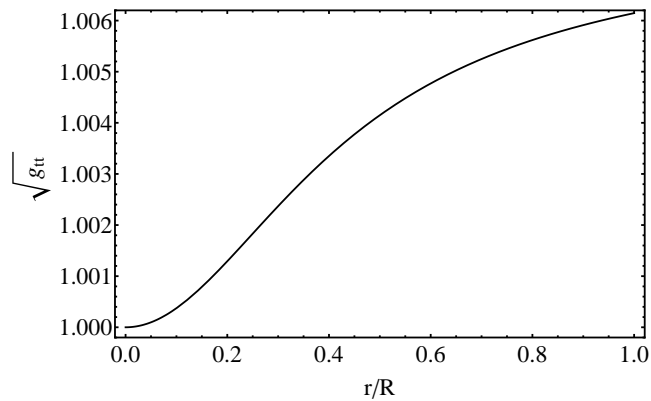}}
\caption{$\sqrt{g_{tt}}$ as a function of the radial distance for a zero temperature white dwarf with mass $M$= 1.44$M_{\odot}$ and radius $R$=1000 km.}
\label{fig:expr1}
\end{figure}

For the sake of simplicity, throughout the paper we use a uniform temperature
profile for isothermal cores of WDs, i.e. WDs without an outer envelop
(atmosphere).  The atmosphere serves as an insulator and its effect on the
structure of WDs can be neglected. In order to justify a constant temperature
profile within the core, we considered the equilibrium condition for rotating
hot relativistic stars, which is given by
$T\sqrt{g_{tt}+2g_{t\phi}\Omega+g_{\phi\phi}\Omega^2}=constant$
\cite{zelnov71}, where $g_{ik}$ are the components of the metric tensor in GR
and $\Omega$ is the angular velocity of a star. For a non-rotating star the
condition reduces to the well-known Tolman condition $T \sqrt{g_{tt}} =
constant$ \cite{tolman1930} , where $T$ is the local temperature.

In the classical limit $\sqrt{g_{tt}} \approx 1-\Phi/c^2$, where $\Phi=
\Phi(r)$ is the internal Newtonian gravitational potential found from
Eq.~(\ref{field}). We constructed $ \sqrt{g_{tt}}$ as a function of $r/R$ for a
WD with mass 1.44$M_{\odot}$ and radius 1000 km in Fig.~\ref{fig:expr1}, as an
example. One can see that the function $\sqrt{g_{tt}}$ changes slightly from
the center to the surface of the isothermal WD core less than 1$\%$. Hence, one
can safely use the classical equilibrium condition $T = constant$ for hot WDs.
This is the foremost argument to adopt the constant temperature profile. For
low mass white dwarfs the function $\sqrt{g_{tt}}$  changes even less than in
the previous case, since when the mass decreases, the radius increases and
$\Phi$ decreases as well. Thus, for the cores of WDs the constant temperature
profile is a sound assumption.

\section{Results and discussion about rotating white dwarfs at finite temperatures}

\begin{figure}
\centerline{\includegraphics[width=8.5cm]{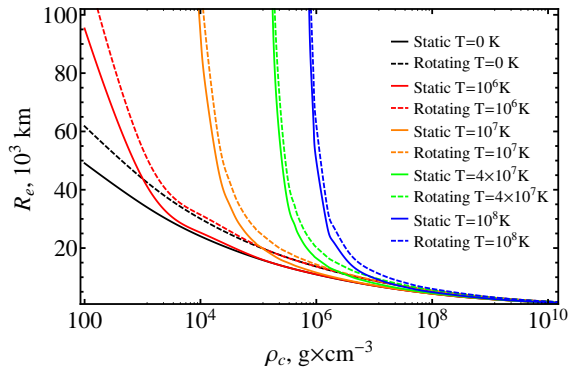}}
\caption{ Radius versus central density (colour online).}\label{fig:Rrho}
\end{figure}

\begin{figure}
\centerline{\includegraphics[width=8.5cm]{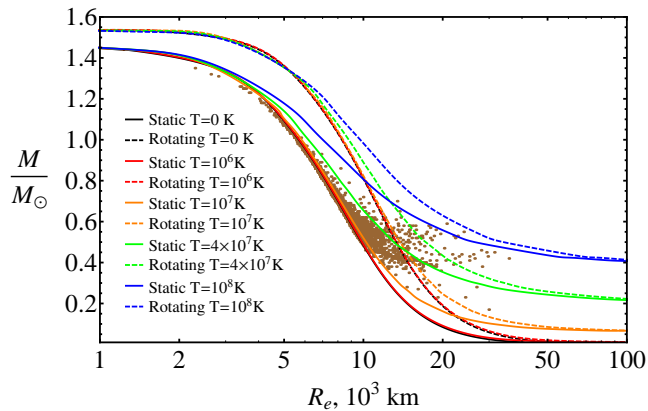}}
\caption{ Mass versus radius (colour online).}\label{fig:MR}
\end{figure}

The Hartle formalism \cite{Hartle, Hartle2, bosh2016} was invoked in classical
physics to calculate the sought parameters of uniformly rotating WDs employing
the Chandrasekhar EoS at finite temperatures. The final results are depicted in
Figs.~\ref{fig:Rrho} and \ref{fig:MR}.
Fig.~\ref{fig:Rrho} shows the equatorial radius as a function of the central
density and temperature for both rotating and static WDs. It is obvious that
hot WDs possess larger radii than cold ones. For increasing central densities,
WDs become more gravitationally bound and spherical. By examining only static
WDs one can  easily calculate the thickness of a hot non-degenerate layer on
top of the cold degenerate one. Consequently, this effect translates also to
rotating WDs.

Fig.~\ref{fig:MR} shows the mass-radius relation for hot static and rotating
WDs superposed over the estimated mass-radius data points from the Sloan
Digital Sky Survey Data Release 4 \cite{Tremblay2011} (brown points). It is
evident that this relation is very different from the degenerate case, in
particular, for small masses and large radii, depending on the temperature of
the isothermal core. The data points are consistent with the theoretical
mass-radius relation.

From the astrophysical context the mass-radius relations for hot WDs play a
pivotal role. As one can see from Fig.~\ref{fig:MR} for the fixed mass the
radius of a WD can be diverse depending on the values of the rotation period
and temperature. From observations, unlike the radius of stars, it is
relatively easy to measure the mass. Therefore, the calculation of radius is a
very delicate problem as the small corrections due to the rotation, GR and the
effects of finite temperatures become more dominant in radius but not in mass
\cite{carvalho2018}.

In Figs.~\ref{fig:Rrho} and \ref{fig:MR} rotating WDs are at the Keplerian
sequence. All realistic uniformly rotating WDs will be in between the static
and mass shedding limit for a fixed temperature. It should be noted, that here
we consider only the temperature of the WD isothermal core $T_c$. The
interrelation of the core temperature and the observed effective surface
temperature  $T_{eff}$ is given via the Koester relation as
$T_{eff}^4/g=2.05\times10^{-10} T_c^{2.56}$, where $g$ is the surface gravity
\cite{koester3}. By employing the Koester formula one can show easily that our
calculations are compatible and consisted with the observational data for WDs
\cite{carvalho2014, Tremblay2011}.

\section{Conclusion}

Mass-radius and radius-central density relations of static and rotating, cold and hot WDs were calculated using the Chandrasekhar EoS. The effects of finite temperatures were accounted for in the EoS. The effects of rotation, such as the deformation of a star, extra mass due to the balance of the centrifugal force and gravity, were investigated within the Hartle formalism in classical physics.

It was shown that in the construction of a realistic model of WDs the effects of finite temperatures and rotation must be accounted for self-consistently. Therefore, unlike in previous studies, here the effects of rotation and finite temperatures were considered together in all our calculations. It was illustrated that for low-mass WDs the effects of temperature are more prominent than for massive WDs. Instead, the rotation affects the structure of WDs in all mass ranges. Consequently, rotation gives an additional degree of freedom for both cold and hot WDs, as expected.

Moreover, we considered the temperatures of the isothermal cores of WDs. For comparison with the observed effective surface temperatures of WDs, the Koester formula must be used, which establishes the interrelation between the temperatures of the atmosphere and the isothermal core of a WD. The mass-radius relations obtained in this work are consistent with observations \cite{Tremblay2011}.

The astrophysical implications of rotating cold and hot WDs are widespread
\cite{boshizzo,rueda2013,boshmg13b,bosh2017,bosh2018}. It is clear that the
inclusion of the magnetic field and nuclear composition will broaden the
applications of WDs to a further extent
\cite{malheiro2012,coelho2014,coelho2014b,lobato2016,coelho2017,alvear2017,alvear2017b,alvear2018}.
Therefore, it would be interesting to continue our research taking into account
the nuclear composition of the WDs matter along with rotation, temperature and
magnetic field. This problem will be considered in our future investigations.

{\bf Acknowledgement}. The author expresses his deep gratitude to the
organizers of the 3rd Zeldovich Meeting for the invitation and for the
organization of the excellent conference. The work was supported in part by
Nazarbayev University Faculty Development Competitive Research Grants: Quantum
gravity from outer space and the search for new extreme astrophysical
phenomena, Grant No. 090118FD5348 and by the MES of the RK, Program IRN:
BR05236494.



\end{document}